# Mathematical Model for the Optimal Utilization Percentile in M/M/1 Systems:

## A Contribution about Knees in Performance Curves

Francisco A. González-Horta, Rogerio A. Enríquez-Caldera, Juan M. Ramírez-Cortés, Jorge Martínez-Carballido
INAOE, Department of Electronics
Luis Enrique Erro #1, Tonantzintla, 72840, Puebla, México. {fglez, rogerio, jmram, jmc}@inaoep.mx

Eldamira Buenfil-Alpuche
Master in Computer Science
Professional Associated to INAOE
eldamira@gmail.com

*Abstract*— Performance curves of queuing systems can be analyzed by separating them into three regions: the flat region, the knee region, and the exponential region. Practical considerations, usually locate the knee region between 70-90% of the theoretical maximum utilization. However, there is not a clear agreement about where the boundaries between regions are, and where exactly the utilization knee is located. An open debate about knees in performance curves was undertaken at least 20 years ago. This historical debate is mainly divided between those who claim that a knee in the curve is not a well-defined term in mathematics, or it is a subjective and not really meaningful concept, and those who define knees mathematically and consider their relevance and application. In this paper, we present a mathematical model and analysis for identifying the three mentioned regions on performance curves for M/M/1 systems; specifically, we found the knees, or optimal utilization percentiles, at the vertices of the hyperbolas that relate response time as a function of utilization. Using these results, we argue that an adaptive and optimal queuing system could be deployed by keeping load and throughput within the knee region.

## I. INTRODUCTION

According to Bose [1] and Kleinrock [2], queuing is a basic phenomenon that arises whenever a shared resource (server) of finite capacity is accessed for service by a large number of jobs or customers. Queues or waiting lines are frequent in many systems and daily life situations; e.g., when you wait for a free ATM to take money out of your bank account, or when a computer inputs data packets into the network, and after a time delay, they are delivered to the destination computer.

Nobody likes to wait too long for a service on a shared resource, thus, one major goal for the performance of a queuing system is to reduce, as much as possible, the system *response time* (R) or delay. However, making a very fast service may have a very high cost. The response time may be reduced if the system capacity is increased, but, if we extend the capacity more than necessary, then the costs for system maintenance and construction will rise, and surely, many resources will be wasted unnecessarily. Therefore, in order to reduce those costs, the system should operate with the greater load or throughput as possible, i.e., allowing to attend as many jobs or customers as possible per unit of time.

One measure of load is *utilization* (U), which is the resource usage divided by resource capacity for a given time interval. Thus, increasing the utilization rate of a queuing system is another relevant goal for raising its performance. However, as utilization for a resource goes up, so does the response time, meanwhile in the opposite way, as the response time goes down, so does the utilization rate. This means that there is a conflict between reducing the response time and increasing the system utilization, both goals cannot be optimized simultaneously, unless we find a compromised or balanced solution. In optimization theory, a Pareto improvement [3] can be made by improving one goal as long as the change that made that goal better off does not make the other goal worse off. When no further Pareto improvements can be made, then the solution is called Pareto optimal or non-dominated solutions. The utilization percentile at which this optimal balance occurs is called the *knee*. This is the point at which load is maximized with minimal negative impact to response times.

A recent article by Cary Millsap [4], about performance for computer software, brought back a discussion on an old debate that has been rounding out the queuing literature for more than 20 years. In 1988, Stephen Samson [5] argued that, at least for M/M/1 queuing systems (i.e., single-server queues where both inter-arrival times and service times follow the exponential distribution), no "knee" appears in their performance curves. Moreover, Samson wrote: "In most cases there is not a knee, no matter how much we wish to find one." Since that moment, a historical debate was initiated between those who support the Samson's claim about knees, e.g., [6][7], and those who argue the existence and relevance of the "knee in the curve" [4]. In this paper, we provide a mathematical approach based on differential calculus to find a "knee" or optimal balance point between conflicting goals involving minimizing response times and maximizing utilization rates for M/M/1 systems. In difference with Millsap [4], who located the knee utilization for a single-server system at U = 50% independently of any other performance parameter, we claim that the knee location is dependent on the average service time for each arrival (S). This knee occurs where the hyperbola, relating U and R, makes its sharpest turn, corresponding to its vertex located at the point $(U, R) = (1 - \sqrt{S}, \sqrt{S})$.

This paper is organized as follows. Section II presents an overview of proposals arguing in favor or against the existence and location of knees in performance curves. Section III starts by presenting a synopsis of the M/M/1 queue model and its nomenclature, then, it shows the mathematical model and analysis for the optimal utilization and the optimal throughput in M/M/1 performance graphs. Thereafter, we propose a region of optimality based on the hyperbola latus rectum. Section IV discusses the relevance of the knee concept and its application to adaptive and optimal communication networks. Finally, Section V concludes the paper with a summary of contributions and future work.

## II. OPEN DEBATE ABOUT KNEES

### A. Proposals Against the Existence of Knees

Neil Gunther in [6] makes a rigorous but unconventional study about knees. He, in fact, analyzed several of the concepts we use in this paper. For instance, he described the hyperbola vertex as an optimum, and he used the endpoints of the latus rectum to find alternative optimum points. Surprisingly, he arrived to the following conclusion: there is no "knee" on the response time curve, even in the case of M/M/1 systems; the same conclusion that Samson arrived in 1998.

A detailed analysis of Gunther's argumentation revealed that he arrived to such conclusion because he analyzed only a normalized response time function (R/S). The R/S function equals to $1/(1-U)$ which corresponds to the R(U) function with S = 1. As we show in this paper, the curve for S = 1 is one of the most inefficient performance curves, because such curve is for an unconventional large number of service times. This is the reason why Gunther, declined his interest in considering the vertex of a hyperbola as the knee in the curve.

Ley [7] reviewed ten different definitions about the knee concept, but, he concluded that there is no a clear definition of what constitutes the knee in the curve, and that all the definitions he collected do not agree with the traditional 70% utilization level. As we prove in this paper, the traditional 70-90% for the optimal utilization range is a myth, because it depends on the service capacity, which makes such a value not a universal constant.

### B. Proposals in Favor of the Existence of Knees

Millsap [4] argues in favor of the existence of knees in performance curves. His paper in fact is quite motivating and provides many useful insights into the fundamentals of performance and further details about this historic debate. He published, in that paper, a table of knee values expressed in utilization percentiles for different number of servers in M/M/m systems. Particularly, for M/M/1 systems he claimed the knee value is 50%. He mentioned that the knee values for an arbitrary number of servers are difficult to calculate, but he also said that the only parameter required to compute them was the number of service channels or servers. However, we disagree on that issue because as we show in this paper, the sharpest point in the curve is dependent on the capacity of the system. We deduced that he obtained such knee values by minimizing the function R/U defined for a specific number of servers. Obtaining the turning points for such function, i.e., making d(R/U)/dU = 0, he calculated the U value of 50% independent of S. The problem with dividing the R function by U is that the R/U function is undefined at U = 0, which is inconsistent with the valid value of R at no load.

## III. MATHEMATICAL MODEL AND ANALYSIS

### A. The M/M/1 Queue Model and Parameters

The simplest queuing system is represented by the Kendall notation as M / M / 1 / ∞ / FCFS. This means that customers arrive according to a Poisson process (first M), they request exponentially distributed service times from the server (second M), the system has only one server, an infinite waiting queue, and customers are served on a First Come First Served (FCFS) basis. For simplicity, this queuing system is sometimes named an M/M/1 system.

As it was neatly described by Chee-Hock in [8], the single-server queue is a place where customers arrive individually to obtain service from a service facility. The service facility contains one server that can serve one customer at a time. If the server is idle, the customer is served immediately. Otherwise, the arriving customer joins a waiting queue. This customer will receive his service later, either when he reaches the head of the waiting queue or according to some *service discipline*. When the server has completed serving a customer, the customer departs. Along this paper, the generic terms 'customers' and 'servers' are in line with queuing literature, but they take various forms in different application domains; e.g., in the case of a data switching network, 'customers' are data packets and 'servers' are the transmission channels.

The M/M/1 system is depicted in Fig. 1. This figure also illustrates some important parameters associated with the queuing model. We describe them briefly.

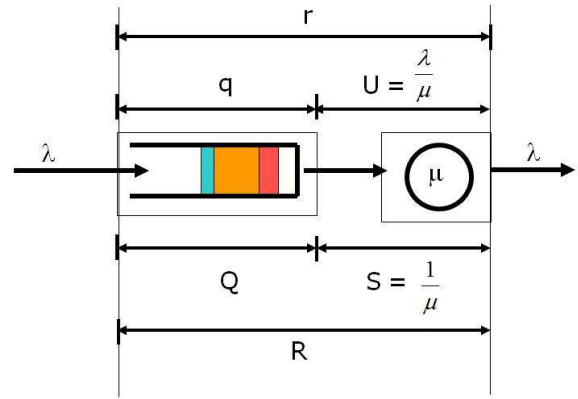

Figure 1. The M/M/1 queue model and parameters at steady state.

- λ, is the average arrival rate or the mean number of customers arriving at the system per unit of time. In

steady state, the rate for arrivals and departures is the same. Moreover, if the waiting line buffer is infinite, then λ also represents the *throughput* of the system, i.e., the mean number of customers that are served in a single unit of time. The domain of this variable is $\lambda \geq 0$.

- μ, is the average service rate or the mean number of customers that are served by the service facility per unit of time. The operating condition $\lambda < \mu$ states the theoretical maximum input rate for the queuing system at $\lambda_{max} = \mu$. If this condition is not achieved (i.e., if $\lambda > \mu$), the number of customers in the waiting line will grow without limit collapsing the system.
- S, is the average service time per customer, it is defined as the reciprocal of μ; i.e., the time interval between the dispatching of a customer to the server and the departure of that customer from the server. The service time cannot be avoided in real scenarios, thus, $S > 0$.
- U is the utilization rate or the fraction of time in which the server is busy. It is obtained as the arrival rate divided by the service rate and it can be expressed as U = λS. For $0 \leq \lambda < \mu$, the domain for U is $0 \leq U < 1$.
- R is the average time that a customer spends in the whole system, waiting and being served; aka, the mean residence time, response time, or delay. $R = 1/(\mu-\lambda)$ or $R = S/(1-U)$, and the domain for R is $S \leq R < \infty$.
- r is the expected number of customers resident in the whole system, including the customers being served (if any) and the customers waiting (if any). This parameter is defined by $r = \lambda/(\mu-\lambda)$ or $r = U/(1-U)$ or $r = R\lambda$. The domain for r is $0 \leq r < \infty$.
- Q is the queuing delay or the mean time that a customer spends in a queue waiting to be serviced. $Q = \lambda/(\mu(\mu-\lambda))$, $Q = R-S$, and the range for Q is $0 \leq Q < \infty$.
- q is the average number of customers waiting in the queue. $q = \lambda Q$, $q = \lambda^2/[\mu(\mu-\lambda)]$, $q = r-U$, with $0 \leq q < \infty$.

In M/M/1 queuing systems, the inter-arrival times and the service times follow the exponential distribution, this means that the arrival and service processes are Poisson (or random). The exponential distribution is the only continuous function that has the *memoryless* (M) property, and thus, it is commonly used to model stochastic processes [11]. Examples of random variables that are well-modeled by the Poisson process are: the number of goals in a soccer match, the number of raindrops falling over an area, the time it takes before your next telephone call, the arrival of customers in a queue, etc.

*B. The Optimal Utilization Percentile (The Knee)*

The performance curves of a queuing system can be obtained by plotting different performance parameters; in particular, we concentrate on relations between R, U, λ, and S or μ. We consider the following relations:

$$R(U; S) = S/(1-U), \text{ for } S > 0 \text{ and } 0 \leq U < 1. \quad (1)$$
$$R(\lambda; \mu) = 1/(\mu - \lambda), \text{ for } \mu > 0 \text{ and } 0 \leq \lambda < \mu. \quad (2)$$

We use (1) for analyzing the relation between response time R and utilization or traffic intensity U, and (2) for studying the relation between delay R and throughput λ. Equations (1) and (2) also show how the response time is a function of service capacity, described by the service time S in (1) and the service rate μ in (2). We will sketch the graphs of these equations for different capacity parameters.

Fig. 2 shows a plot for R(U; S) illustrating the behavior of response time as a function of utilization. Here, each curve is plotted for different values of S, S = 2 (black), 1, 1/2, 1/4, 1/8, 1/16 (blue). Notice the aspect ratio in the plot is 1-to-1, to avoid what Gunther [9] calls an "optical illusion" produced by using different aspect ratios in the utilization and response time axes, which might result in a misconception about the utilization knees. Notice also that, as Gunther showed in [6], the graph for the response time function can be depicted for an extended range of utilization values $-\infty < U < \infty$, even if it does not make physical sense; therefore, we demarcate at Fig. 2 the actual service utilization range $0 \leq U < 1$ as the region of meaningful performance metrics in between bold blue lines.

Fig. 2 stresses the hyperbolic characteristic of the response time function by extending the utilization range. The gray dotted line highlights the transversal axis of the hyperbolae and pinpoints its sharpest turns or vertices. Notice how each curve becomes sharper as S diminishes or the system capacity increases; however, when $S > 1$, the vertices jump to the negative utilization region and response times within the performance region of interest (blue lines) fluctuate rapidly with small changes of load. It is important to design queuing systems operating with S values lower than 1, whatever the time unit is chosen, e.g., seconds, minutes, hours, etc.

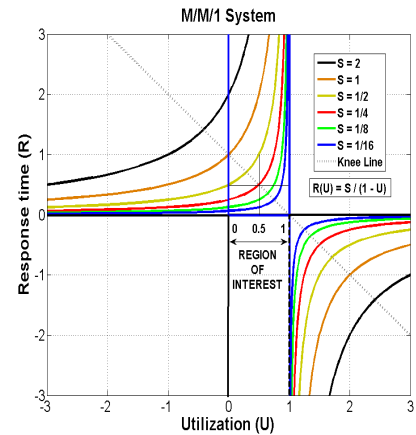

Figure 2. Plot of response time vs utilization for distinct service times.

*1) Hyperbola Vertex as an Optimum.* Considering the optimality condition that indicates a balance between changes in U and changes in R, the point where a differential increment of U yields the same increment of R, or dR = dU, is the point where the rate of change of R with respect to U (dR/dU) is equal to 1. This condition is reached exactly at the vertex of a hyperbola. In fact, that is the point that we use to divide the R(U) function initially in three sections: a) the flat section, where the gradient of the curve is dR/dU < 1, indicating a small but constant increment in response time at low load; b) the knee or optimal utilization point, located at the hyperbola vertex, where dR/dU = 1; and c) the exponential section, where dR/dU > 1, indicating that response time rises exponentially at high load.

Our performance goals are maximizing load and minimizing response time. According to Pareto optimality, starting at no load (U = 0), we make improvements or increments in load as long as the resulting increments in response times do not exceed the increments in load (i.e., dR < dU). Thus, the point where optimal balance occurs is the vertex of the hyperbola. Differentiating equation (1) and making $R'(U) = 1$, we obtain the coordinates for the vertices at $(U, R) = (1 - \sqrt{S}, \sqrt{S})$. Notice that the optimal utilization percentile is a function of S, and depending on the service capacity, this optimum occur at U = 0.5, only if S = 1/4; at U > 0.5 (the high-load zone), if S < 1/4; at U < 0.5 (the low-load zone), if S > 1/4; or even at U = 0 if S = 1.

*2) Latus Rectum as an Optimal Region.* Practical considerations such as response time requirements or buffer sizes, are usually interested in a region of optimal utilization rather than a single optimal point (knee) in the curve. This optimal region is usually located between 70-90% of the theoretical maximum utilization. However, there is no agreement on defining this knee region. We argue that the *latus rectum* of the hyperbola can be used to establish this region of optimality.

Consider Fig. 3. The *Latus Rectum* [10] is the line segment passing through a focus of a hyperbola, which is perpendicular to the transversal axis and has both endpoints (P, Q) on the intersection with the curve (P, Q are indicated just for the curve S = 1).

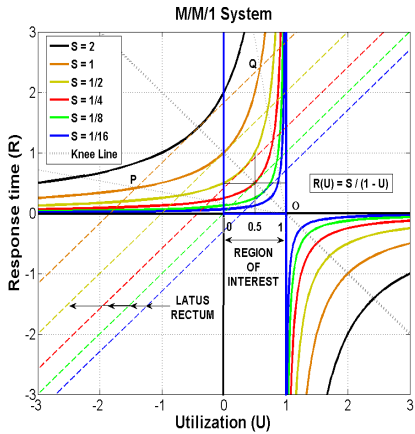

Figure 3. Plot of R(U) for distinct service times (S) with Latus Rectum.

The dashed lines with slope 1 in Fig. 3, represent the graphs of the latus rectum for each sketched hyperbola with S ≤ 1. It can be observed that different scenarios may occur in relation to where the endpoints of the latus rectum are with respect to the region of interest: a) both endpoints are within the bold blue lines, b) one endpoint is in and the other is out, and c) both endpoints are out of the blue lines. The selection of one of these scenarios is, again, dependent on the value of S.

*3) Boundaries of the Knee Region*

The following analysis is intended to obtain the coordinates for both endpoints of the latus rectum and its length in a general way, such that, it can be obtained for any valid value of S.

We already obtained the coordinates for the vertices, now we obtain the coordinates for the foci intersecting the latus rectum of each hyperbola. The distance from the center $O$ (1,0) to the vertex $V(1 - \sqrt{S}, \sqrt{S})$ of the hyperbola is $OV = \sqrt{2S}$. Using this value, we calculate the distance from center to focus as $OF = 2\sqrt{S}$. Thus, the coordinates for the foci are $F(1 - \sqrt{2S}, \sqrt{2S})$. The equation of the straight line representing the latus rectum is: $R = U - 1 + 2\sqrt{2S}$. Now, intersecting the latus rectum equation with the response time function, we obtain the coordinates for the latus rectum endpoints: $P(1 - \sqrt{2S} - \sqrt{S}, \sqrt{2S} - \sqrt{S})$ and $Q(1 - \sqrt{2S} + \sqrt{S}, \sqrt{2S} + \sqrt{S})$. Thus, the boundaries for the three regions that can be used to study the response time versus utilization performance curves can be stated as: a) *flat region*, the locus of points before P, b) *knee region*, the locus of points between P and Q, and c) *exponential region*, the locus of points after Q. Using the distance between two points we obtain the length of latus rectum as: $2\sqrt{2S}$.

Considering that the region of interest for R(U) is 0 ≤ U < 1 and its knee region is $\left(1 - \sqrt{2S} - \sqrt{S}\right) \leq U \leq (1 - \sqrt{2S} + \sqrt{S})$, we calculate the service capacity condition that makes both endpoints of the knee region reside in the interest region; this condition is $0 < S \leq 3 - 2\sqrt{2}$.

*4) Knees in Delay-Throughput Curves*

Considering (2), we analyze how the response time R (delay), arrival rate λ (throughput), service capacity μ (bandwidth), and delay knees (hyperbola vertices), are all related to system performance. Fig. 4 shows a plot which we use to illustrate and explain such relation. Here, each hyperbola is plotted for different values of μ, μ = 1 (brown), 2, 4, 8, 16 (blue). Notice again, how the vertices V of the hyperbolas occur precisely when $dR/d\lambda = 1$ and they represent the sharpest points in the curve. Using this condition we locate the vertices of the upper-half hyperbolae at $V(\mu - 1, 1)$. The coordinates for the centers are $O(\mu, 0)$. With the coordinates for O and V, obtain the distance to the vertices as $OV = \sqrt{2}$. Using the distance to the vertex and the definition of hyperbola, we obtain the distance to the focus as $OF = 2$. Knowing the distance to the focus, we obtain the coordinates of the focus at $F(\mu - \sqrt{2}, \sqrt{2})$. Using the coordinates for F and definition of latus rectum, we obtain the equation for latus rectum as: $R = \lambda - \mu + \sqrt{8}$. Intersecting the latus rectum equation with the hyperbola, we obtain the coordinates for the endpoints of the latus rectum at $P: (\mu - \sqrt{2} - 1, \sqrt{2} - 1)$ and $Q: (\mu - \sqrt{2} + 1, \sqrt{2} + 1)$. Thus, the length of latus rectum is $\sqrt{8}$.

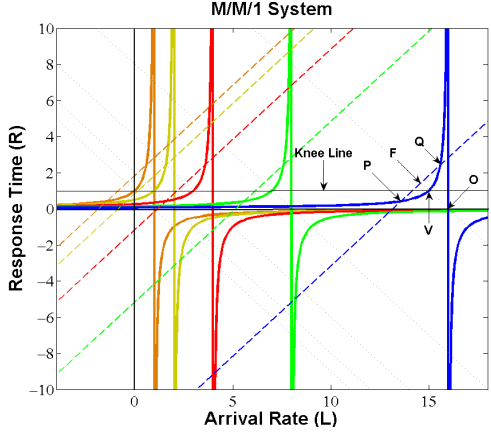

Figure 4. Plot of R(λ) for distinct service rates (μ) with Latus Rectum.

The region of interest is $0 \leq \lambda < \mu$. Notice that for low capacity systems, the knee region: $(\mu - \sqrt{2} - 1) \leq \lambda \leq (\mu - \sqrt{2} + 1)$ may have its lower endpoint outside the region of interest. In order to have both endpoints of the latus rectum within the region of interest, the following condition must be satisfied: $\mu \geq \sqrt{2} + 1$.

System performance is measured by delay R (i.e., the time a customer stays within the system), and by throughput λ (i.e., the number of customers per unit of time that can pass through the system). Throughput is a measure of the system capacity. Delay and throughput are closely related by (2), as throughput approaches 100% of service capacity or bandwidth (μ), delay increases rapidly.

*5) Summary of Results*

Table I shows a summary of major results obtained from the analysis presented in this paper.

TABLE I. SUMMARY OF MAJOR RESULTS

| Useful Results | Response Times for Load and Throughput | |
|---|---|---|
| | $R(U) = S/(1-U)$ | $R(\lambda) = 1/(\mu - \lambda)$ |
| Region of interest | $S > 0, 0 \leq U < 1$ | $\mu > 0, 0 \leq \lambda < \mu$ |
| Knee values | $U = 1 - \sqrt{S}$ | $\lambda = \mu - 1$ |
| Knee coordinates | $(U, R(U)): (1 - \sqrt{S}, \sqrt{S})$ | $(\lambda, R(\lambda)): (\mu - 1, 1)$ |
| Knee regions | $U \geq 1 - \sqrt{2S} - \sqrt{S}$ <br> $U \leq 1 - \sqrt{2S} + \sqrt{S}$ | $\lambda \geq \mu - \sqrt{2} - 1$ <br> $\lambda \leq \mu - \sqrt{2} + 1$ |
| Latus rectum endpoints | $P: (1 - \sqrt{2S} - \sqrt{S}, \sqrt{2S} - \sqrt{S})$ <br> $Q: (1 - \sqrt{2S} + \sqrt{S}, \sqrt{2S} + \sqrt{S})$ | $P: (\mu - \sqrt{2} - 1, \sqrt{2} - 1)$ <br> $Q: (\mu - \sqrt{2} + 1, \sqrt{2} + 1)$ |
| Length of latus rectum | $2\sqrt{2S}$ | $2\sqrt{2}$ |
| Required service capacity | $0 < S \leq 3 - 2\sqrt{2}$ | $\mu \geq \sqrt{2} + 1$ |

These closed-form results show that there is a knee or optimal value for load and throughput in simple M/M/1 queuing systems. Moreover, based on the latus rectum endpoints it is possible to define optimality regions for load and throughput. It can also be observed that both, the knee values and knee regions are dependent on the service capacity of the queuing system. Therefore, it was important for us to determine what would be the required service capacity that will make the knee regions to operate within the regions of interest. The service capacity can be increased by reducing the service time S or by increasing the number of servers. Although we only analyzed the case of increasing the service capacity by reducing S, similar results can be obtained if the case of increasing the number of servers is considered.

## IV. RELEVANCE AND APPLICATION OF THE KNEE

### A. The Relevance of the Knee Region

Why is the knee value so important? The answer to this question is related to the consequences of having a system operating outside its optimal region. On one hand, if the system operates at the flat region, it is likely that the system capacity is oversized, and therefore, many valuable resources may be wasted. On the other hand, if the system operates at the exponential region, it is likely that the system capacity is undersized, and hence, response times will fluctuate severely even with microscopic changes in load or throughput. The system operation in the exponential region may take the system to instability, oscillating congestion, severe delays, or the worst scenario, to a collapse. Hence, overall on systems with *random arrivals*, it is vital to manage load and throughput so that they do not operate outside the knee region.

At M/M/1 system queues, we do not know *exactly* when the next arrival request or service request is coming; therefore, arrivals have a non-deterministic or random behavior. The M/M/1 model considers an exponential probability distribution for arrivals and service requests. The exponential distribution assumes a higher probability for small inter arrival times. This implies that arrivals will tend to cluster and cause temporary spikes in utilization, as it was mentioned by Millsap [4]. Temporary spikes in utilization beyond the *knee* region may cause serious performance problems or quality of service (QoS) degradation if they exceed a few seconds in duration. This is the reason the knee region is so important on a system with random arrivals.

Once we mentioned the consequences of a system with oversized or undersized capacity, the question is how to determine the adequate capacity that makes a system to operate within its optimal region? *Capacity management* or *capacity planning* is a task intended to answer that question. A first consideration about the service capacity of a queuing system is that it should be calculated so that the optimal region lay down within the region of interest. The last row in Table I shows the conditions for S and μ that make the knee regions for R(U) and R(λ) to be within the region of interest. A second consideration is to estimate the service capacity according to specific service quality requirements; i.e., the expected amount of traffic intensity, throughput, and response times, particularly at peak times. The knee regions can be computed to meet these QoS requirements. If service capacity cannot be changed dynamically, then the system would operate within its optimal region at peak times, but it would operate outside its optimal region at different load conditions than the peak times. On the contrary, if capacity could be changed dynamically, then the knee regions could change accordingly so that the system operates always within its optimal region. This implies to change *adaptively* the knee regions according to different load conditions (at low load, at peak times, or at excessive high loads). In this way, we believe that and adaptive and optimal queuing system could be deployed by dynamically managing the service capacity in order to keep load and throughput within their knee regions. Therefore, the knowledge about knees is fundamental for capacity management.

### B. The Knee Concept Applied to Communication Networks

The M/M/1/∞/FCFS model discussed earlier is simple and useful if we just want to have a first estimation of a system's performance. However, as Chee-Hoc indicates in [8], it becomes a bit unrealistic when it is applied to real-life problems, where they often have a finite waiting queue instead of one that can accommodate an infinite number of customers. A single isolated M/M/1 model may have certain limitations to represent real-life complex queuing systems; however, the networks of queues, whereby the departures of some queues feed into other queues, are a more realistic model for a system with many shared resources interacting with each other. In this way, a model for a virtual circuit in packet switching networks can be designed in terms of a network of tandem M/M/1 queues. Therefore, we can say that M/M/1 queues are the building blocks for all the queuing theory, as we will show in the next discussion.

*1) Using the M/M/1 model:*

In order to apply the knee concept to a real-life queuing system, we now immerse a bit into the field of communication networks. Two fundamental performance measures of a network are: delay (D) and throughput (T). First notice that we changed the symbols for the equivalent terms we used before as R for delay and λ for throughput in the M/M/1 model. The term delay (or end-to-end delay) specifies how long it takes for a single bit of data to travel across the network cloud from source to destination, as Fig. 5 illustrates. Comer in [12] cleverly describes four types of delays that may occur within the network: propagation delays, switching delays, access delays and queuing delays. All these types of delays contribute to the total delay, which is measured in seconds or fractions of seconds. The other fundamental performance parameter of a network is throughput. Throughput is the rate at which bits of data can pass through the network, and is usually specified in bits per second. The rate at which the hardware can transfer bits is called bandwidth (B). It is impossible for a user to send data faster than the bandwidth; therefore, bandwidth gives an upper bound on throughput or $0 \leq T < B$. Notice that B is equivalent to the μ parameter defined earlier.

In Fig. 5, we also made a distinction between the rate of bits entering the network from source (γ) and the rate of bits leaving the network at destination (T). Notice that for a stable and lossless network, T = γ; however, if T < γ then there will be a steady build-up of bits within the network and it will eventually become unstable. In case of T > γ, new bits might be being created within the network.

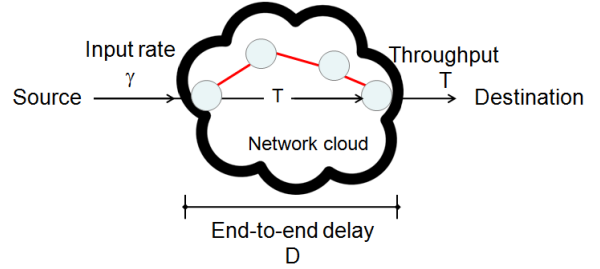

Figure 5. Store and forward data network. Packets traverse from source to destination. Circles are packet switches and red lines are links.

The capacity of a network refers to the maximum number of bits that can simultaneously reside within the network, inclusive of those bits waiting for a shared resource and those that are actually traversing across a link. If $D_0$ denotes the average delay when the network is idle or when the current utilization (U) is 0, then the network capacity is defined by the Little's Law as D × T (delay-throughput product). According to [8], Little's Law can be applied to all types of queuing systems, including priority queuing and multi-server systems, which is the case for packet-switching networks. Notice that $D_0$ corresponds to S, U = T/B corresponds to U = λ/μ, and D × T corresponds to r = Rλ in the M/M/1 model. When it is desired to measure the capacity of the underlying hardware, the delay-throughput product becomes the delay-bandwidth product.

Congestion in the network occurs when the offered traffic load from the user to the network produces an excess of bits residing within the network, which exceed the design limit. Data entering a congested network will experience longer delays than data entering an idle network. The relationship between delay and congestion is estimated from the current percentage of the network capacity being used, according to the following expression:

$$D = \frac{D_0}{(1-U)}, for\ D_0 > 0\ and\ 0 \leq U < 1. \quad (3)$$

As traffic approaches the network capacity (i.e., as U becomes close to 1), the delay approaches infinity. We already plotted this expression in Figure 2 and Figure 3 and obtained some significant insights summarized in Table I.

*2) Towards an Adaptive-Optimal Congestion Control:*

A simple mechanism for preventing congestion is flow control, which involves regulating the input rate of traffic into the network so that the receiver controls the rate at which it receives data. The main goal of a flow control mechanism is to preserve the levels of throughput and delay within a range of acceptable values. There are basically two types of flow control mechanisms, open-loop and closed-loop. The open-loop flow control is characterized by having no feedback between the source and destination. This mechanism allocates networks resources for a specific traffic flow with necessary previous reservation. Often this type of mechanism is inefficient and results in over-allocation of resources, it also lacks of any adaptability. On the contrary, the closed-loop flow control is characterized by the ability of the network to report pending network congestion back to the source node. By using this feedback, the source can adapt its transmission rate to a lower rate or a higher rate depending on the network conditions. Real protocols that implement different mechanisms of closed-loop flow control are TCP (Transmission Control Protocol) and ABR (Available Bit Rate). On one hand, TCP adaptively increments or decrements the size of a window in order to speed-up or slow-down the transmission rate. On the other hand, ABR uses congestion information generated in the destination and updated at each packet switch on the path to the source, to reduce or increment the transmission rate accordingly.

Congestion is usually caused by unpredictable events. Although the daily peak hour is semi-predictable, congestion can also be random due to breakdowns, insertions, or changes in the network topology. Therefore, an adaptive mechanism to control the traffic flow is imperative to alleviate congestion problems and preserve the stability of the network.

To know where exactly the borders of the optimal load/throughput region are, as identified in this paper, could be used by the end nodes to regulate and control the stability of the network. If the network capacity could be dynamically managed by the source and destination nodes, then new optimal knee regions would be created, giving the possibility that end stations transmit data flows optimally, i.e. minimizing delays and maximizing throughput [11].

The relevance of the results obtained in this paper is that they can be applied to both, the simplest queuing systems (M/M/1) and the most complex communication network. This is because the M/M/1 model is the building block of any complex queuing system and because Little's Law can be applied neatly to all types of queuing systems.

## V. Conclusion and Future Work

This paper presented a mathematical approach to determine the optimal utilization region and the optimal throughput region for M/M/1 queuing systems. It showed that performance curves for such systems can be analyzed by separating them into three regions: the flat region, the knee region, and the exponential region. The mathematical definition of boundaries between these regions is a problem that has not been properly addressed in the literature. The paper showed that this problem has historical roots and it is still an open debate. The major contribution of this paper is the calculation of knees values and knees regions in performance curves for load R(U) and throughput R(λ). The relevance of knees and their applications was discussed showing the consequences of operating a system outside of its optimality region.

Although the knee model proposed in this paper seems to be mathematically consistent, it is still necessary to create tests, in real applications or simulations, which can help us to validate or invalidate this model. Therefore, our future work will be focused on this task.

## Acknowledgment

F.A. González-Horta is a Ph.D. candidate at INAOE. He thanks the financial support received from the National Council of Science and Technology (CONACYT) Mexico, through the doctoral scholarship 58024.